\documentstyle{mn}
\input{psfig}
\newcommand{\etal}{{et al.~}}

\newcommand{\kms}{\>{\rm km}\,{\rm s}^{-1}}

\newcommand{\Mpc}{\>{\rm Mpc}}
\newcommand{\kpc}{\>{\rm kpc}}
\newcommand{\Msun}{\>{\rm M_{\odot}}}
\newcommand{\Lsun}{\>{\rm L_{\odot}}}

\newcommand{\beq}{\begin{equation}}
\newcommand{\eeq}{\end{equation}}

\newcommand{\R}{{\cal R}}

\def\gtsima{$\; \buildrel > \over \sim \;$}
\def\ltsima{$\; \buildrel < \over \sim \;$}
\def\prosima{$\; \buildrel \propto \over \sim \;$}
\def\gsim{\lower.7ex\hbox{\gtsima}}
\def\lsim{\lower.7ex\hbox{\ltsima}}
\def\simgt{\lower.7ex\hbox{\gtsima}}
\def\simlt{\lower.7ex\hbox{\ltsima}}
\def\simpr{\lower.7ex\hbox{\prosima}}

\newcommand{\apj}{ApJ}
\newcommand{\apjs}{ApJS}
\newcommand{\aj}{AJ}
\newcommand{\mnras}{MNRAS}
\newcommand{\aap}{A\&A}

\newcommand{\nat}{Nature}

\newdimen\hssize
\newdimen\hdsize
\begin{document}


\title[The Phase-Space Parameters of Brightest Halo Galaxies]
      {The Phase-Space Parameters of Brightest Halo Galaxies}
\author[F.C. van den Bosch et al.]
       {Frank C. van den Bosch$^{1}$ \thanks{E-mail:vdbosch@phys.ethz.ch}, 
        Simone M. Weinmann$^{1}$, Xiaohu Yang$^{2}$, H.J. Mo$^{2}$, 
\newauthor Cheng Li$^{3}$ and Y.P. Jing$^{4}$\\
        $^1$Department of Physics, Swiss Federal Institute of
         Technology, ETH H\"onggerberg, CH-8093, Zurich,
         Switzerland\\ 
        $^2$Department of Astronomy, University of Massachussets, 710
         North Pleasant Street, Amherst MA 01003-9305, USA\\
        $^3$Center for Astrophysics, University of Science and
         Technology of China, Hefei, 230026, China\\
        $^4$Shanghai Astronomical Observatory; the Partner Group of
         MPA, Nandan Road 80, Shanghai 200030, China}


\date{}

\pagerange{\pageref{firstpage}--\pageref{lastpage}}
\pubyear{2000}

\maketitle

\label{firstpage}


\begin{abstract}
  The brightest galaxy in a dark  matter halo is expected to reside at
  rest at the center of the  halo. In this paper we test this `Central
  Galaxy   Paradigm'  using  group   catalogues  extracted   from  the
  Two-Degree  Field  Galaxy Redshift  Survey  (2dFGRS)  and the  Sloan
  Digital Sky  Survey (SDSS).  For  each group we compute  a parameter
  ${\cal R}$, which is defined  as the difference between the velocity
  of the brightest group galaxy  and the average velocity of the other
  group  members (hereafter  satellites), normalized  by  the unbiased
  estimator  of the  velocity dispersion  of the  satellite  galaxies. 
  Since the  redshift surveys suffer from  incompleteness effects, and
  the  group selection  criterion unavoidably  selects  interlopers, a
  proper comparison  between data  and model needs  to take  this into
  account.   To  this extent  we  use  detailed  mock galaxy  redshift
  surveys, which  are analyzed  in exactly the  same way as  the data,
  thus  allowing for  a fair  comparison.   We show  that the  central
  galaxy paradigm  is inconsistent with  the data at  high confidence,
  and that instead the brightest halo galaxies have a specific kinetic
  energy  that is about  25 percent  of that  of the  satellites. This
  indicates that either central galaxies  reside at the minimum of the
  dark matter  potential, but  that the halo  itself is not  yet fully
  relaxed, or, that  the halo is relaxed, but  that the central galaxy
  oscillates in its potential well.  The former is consistent with the
  fact that  we find a weak  hint that the velocity  bias of brightest
  halo galaxies is larger in more massive haloes, while the latter may
  be indicative of  cored, rather than cusped, dark  matter haloes. We
  discuss  several  implications  of  these findings,  including  mass
  estimates  based  on   satellite  kinematics,  strong  gravitational
  lensing, halo occupation models,  and the frequency and longevity of
  lopsidedness in disk galaxies.
\end{abstract}




\section{Introduction}
\label{sec:intro}

In the  standard picture  of galaxy formation,  hot gas  in virialized
dark  matter  haloes  cools  and  accumulates at  the  center  of  the
potential well, where  it forms a galaxy (White  \& Rees 1978). During
the hierarchical build up of larger and larger structures, haloes with
their  `central' galaxies are  accumulated by  even larger  haloes. At
that point the halo becomes  a subhalo, and the central galaxy becomes
a satellite galaxy.  In the  standard picture, it is envisioned that a
satellite galaxy  no longer  accretes hot gas,  which instead  is only
accreted  by the galaxy  in the  center of  the potential  well (e.g.,
Kauffmann, White \& Guiderdoni  1993; Somerville \& Primack 1999; Cole
\etal 2000).   Since this central galaxy therefore  continues to grow,
it is  expected to be  the brightest, most  massive galaxy in a  halo. 
This  is further assured  by the  fact that  any other  massive galaxy
would quickly  sink to the center  of the potential  well by dynamical
friction to merge with the central galaxy, thus producing an even more
massive  central   galaxy.   Therefore,  according   to  the  standard
paradigm, the  brightest galaxy in a  halo will reside at  rest at the
center of the potential well.  Note that this is clearly a statistical
statement, as it does not  necessarily hold for each individual system
(e.g., non-virialized,  strongly interacting systems).   Hereafter, we
will refer  to this paradigm  as the `Central Galaxy  Paradigm' (CGP),
and use the terms `central galaxy' and `brightest halo galaxy' without
distinction.

The CGP plays an important  role in various areas of astrophysics. For
example,  attempts  to measure  halo  masses  from  the kinematics  of
satellite galaxies,  are always based  on the general  assumption that
the `host' galaxy  is located at rest at the center  of a relaxed halo
(e.g.,  Zaritsky \etal  1993, McKay  \etal 2002;  Brainerd  \& Specian
2003; Prada \etal 2003; van den Bosch \etal 2004).  This assumption is
also used in virtually all mass models of strong gravitational lenses.
On  the   other  hand,  the   observed  frequency  and   longevity  of
lopsidedness in disk galaxies (e.g., Richter \& Sancisi 1994; Zaritsky
\& Rix  1997) is  often interpreted as  evidence for an  actual offset
between galaxy  and halo (e.g.,  Levine \& Sparke 1998).   The central
galaxy paradigm also  plays a role in halo  occupation modeling, where
assumptions  have to  be made  regarding the  spatial  distribution of
galaxies in  haloes in order to compute  the galaxy-galaxy correlation
function  on small scales  (e.g., Scoccimarro  \etal 2001;  Berlind \&
Weinberg 2002; Yang, Mo \& van  den Bosch 2003; van den Bosch, Yang \&
Mo  2003; Magliocchetti \&  Porciani 2003;  Tinker \etal  2004; Zehavi
\etal  2004; Zheng  \etal  2004).  A  statistic  that is  particularly
sensitive to whether the brightest  halo galaxies reside at the center
or  not  is the  cross  correlation  between  dark matter  haloes  and
galaxies (see Yang \etal 2005, in preparation).

This special  dynamical status of the  brightest galaxy in  a halo has
been tested for  the special class of cD  galaxies. Jones \etal (1979)
have  shown that  cDs are  located at  the peak  of the  cluster X-ray
emission,  while Quintana  \&  Lawrie (1982)  used  the kinematics  of
cluster galaxies  to argue that  cDs are at  rest with respect  to the
cluster.   Although this  is in  agreement with  the CGP,  more recent
studies  have revealed  various cases  in which  the cD  galaxy  has a
significant peculiar velocity with respect to the mean velocity of the
other cluster members (e.g., Sharples,  Ellis \& Gray 1988; Hill \etal
1988; Zabludoff, Huchra  \& Geller 1990; Oegerle \&  Hill 1994, 2001). 
Applying a similar study to a dozen poor groups, Zabludoff \& Mulchaey
(1998) and Muchaey \& Zabludoff (1998)  found that the position of the
brightest galaxy in  each group is indistinguishable from  that of the
group  center  or  from the  center  of  the  X-ray emission.  To  our
knowledge, however, the central  galaxy paradigm has never been tested
for a statistically significant sample of dark matter haloes that span
a wide range in masses.  In this paper we use data from the Two-Degree
Field  Galaxy Redshift  Survey (2dFGRS,  Colless \etal  2001)  and the
Sloan  Digital Sky  Survey (SDSS;  York \etal  2000) to  directly test
whether the  brightest galaxies in  dark matter haloes are  located at
rest at  the center of their  potential well.  We  show that, although
the brightest halo galaxies are clearly segregated with respect to the
other galaxies in the same  halo, they have a typical specific kinetic
energy  that is  about  $\sim 25$  percent  of that  of the  satellite
galaxies, and that the CGP is ruled out at a high level of confidence.

This  paper is  organized  as follows.   In Section~\ref{sec:sign}  we
present a statistic  that can be used to test the  CGP, which we apply
to    the   2dFGRS   and    SDSS   in    Section~\ref{sec:data}.    In
Section~\ref{sec:velbias} we describe a  simple model for the velocity
and  spatial bias  of the  brightest halo  galaxies, which  we  use in
Section~\ref{sec:mock}  to  construct  detailed mock  galaxy  redshift
surveys of the 2dFGRS. In Section~\ref{sec:res} we compare these mocks
with  the data  in order  to constrain  the phase-space  parameters of
brightest  halo  galaxies.   Section~\ref{sec:disc} discusses  various
implications  of our  results,  and we  summarize  our conclusions  in
Section~\ref{sec:concl}.

\section{Dynamical Signature of Central Galaxies}
\label{sec:sign}

Observationally, the  only kinematic information that  is available to
test  the central  galaxy  paradigm are  the line-of-sight  velocities
obtained from redshifts. In what follows  we use $v_c$ to refer to the
line-of-sight velocity of the brightest  halo galaxy, and $v_i$ is the
line-of-sight  velocity  of the  $i^{\rm  th}$  satellite galaxy.   In
addition we define the difference $\Delta V = \bar{v}_s - v_c$ between
the {\it  mean} velocity  of the satellite  galaxies ($\bar{v}_s  = {1
  \over N_s}  \sum_{i=1}^{N_s} v_i$) and  that of the central  galaxy. 
If the CGP  is correct and $v_i$ follows  a Gaussian distribution with
velocity dispersion $\sigma_s$, the probability that a halo with $N_s$
satellite galaxies has a value of $\Delta V$ is given by
\begin{equation}
\label{probV}
P(\Delta V) {\rm d}\Delta V = {1 \over \sqrt{2 \pi} \sigma} 
\exp\left[- {(\Delta V)^2 \over 2 \sigma^2}\right] \, {\rm d}\Delta V\,,
\end{equation}
with  $\sigma =  \sigma_s/\sqrt{N_s}$.  Therefore,  in  principle, one
could define the parameter
\begin{equation}
\label{normr}
R = {\sqrt{N_s} (\bar{v}_{s} - v_c) \over \sigma_s}\,,
\end{equation}
and test the CGP by checking whether $R$ follows a normal distribution
with zero  mean and unit  variance.  However, the  velocity dispersion
$\sigma_s$  is generally  unknown, and we have to use its unbiased
estimator
\begin{equation}
\label{sigest}
\hat{\sigma}_s = \sqrt{{1 \over N_s-1} \sum_{i=1}^{N_s} (v_i - \bar{v}_s)^2}
\end{equation}
instead. This allows us to define the modified parameter
\begin{equation}
\label{calr}
\R = {\sqrt{N_s} (\bar{v}_{s} - v_c) \over \hat{\sigma}_s}\,.
\end{equation}
If the  null-hypothesis of  the CGP is  correct, $\R$ should  follow a
Student  $t$-distribution with $\nu=N_s-1$  degrees of  freedom.  Note
that $P_{\nu}(\R)$ approaches a normal distribution with zero mean and
unit variance in the limit $N_s \rightarrow \infty$.
\begin{figure*}
\centerline{\psfig{figure=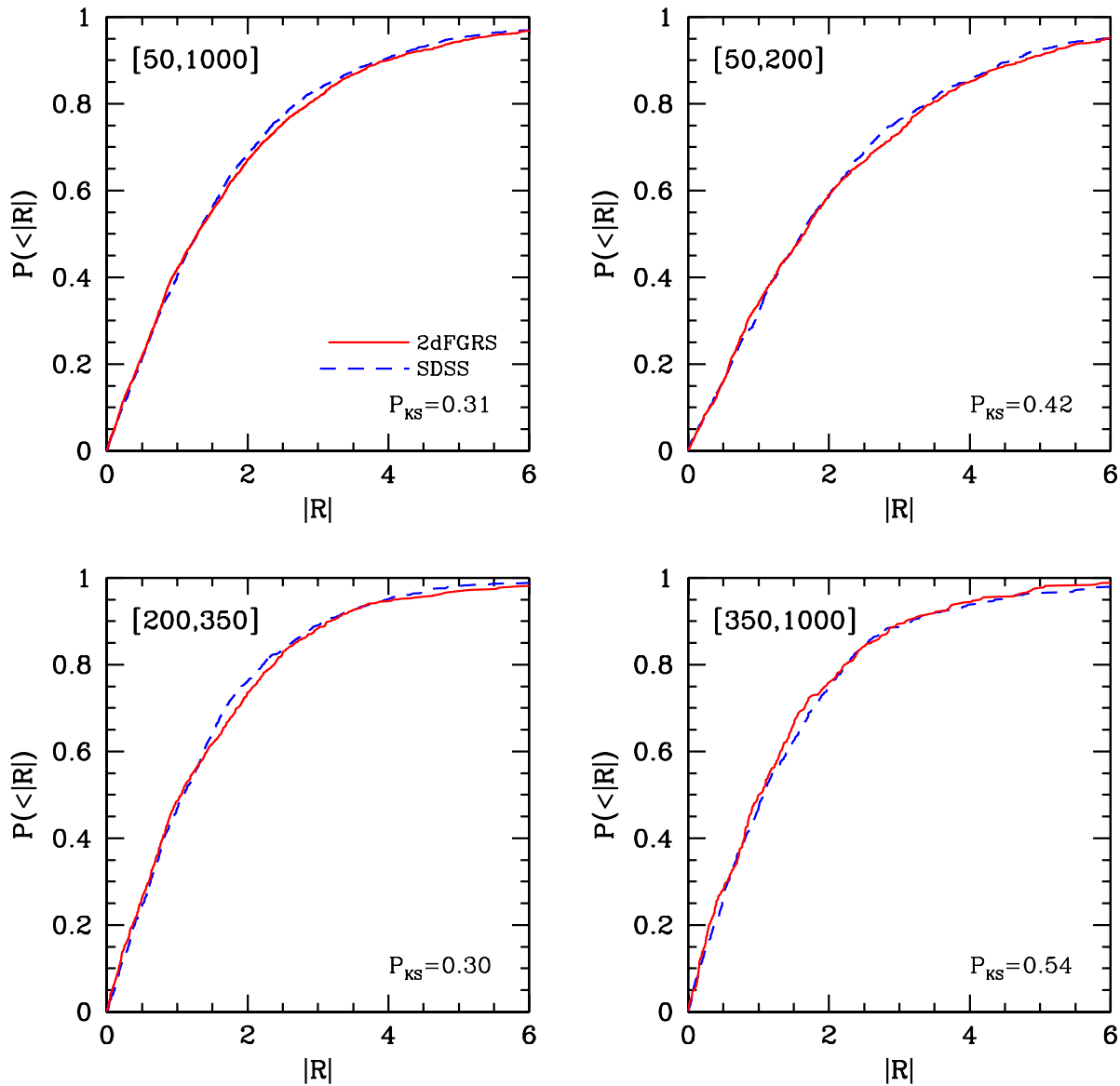,width=0.9\hdsize}}
\caption{A comparison of the cumulative distributions of $\vert {\cal
    R}  \vert$ of  2dFGRS (red,  solid lines)  and SDSS  (blue, dashed
  lines)   groups.   Results   are   shown  for   four  intervals   in
  $\hat{\sigma}_s$, indicated in square brackets in each panel. The KS
  probability, $P_{\rm  KS}$, that  both distributions are  drawn from
  the  same distribution  is also  indicated. Note  that  the $P(\vert
  {\cal R}  \vert)$ from  2dFGRS and SDSS  are in  excellent agreement
  with each other.}
\label{fig:comparison}
\end{figure*}

The  applicability of  this  `$\R$-test' is  strongly  related to  the
ability to  find those  galaxies that belong  to the same  dark matter
halo.   To this  extent  we  use the  halo-based  galaxy group  finder
developed by  Yang \etal  (2005a), which has  been optimized  for this
task. Although this group finder  is well tested and calibrated, it is
not perfect.   In particular, because of redshift  errors and redshift
space  distortions, it  is  unavoidable that  one selects  interlopers
(galaxies that are not associated with the same halo). The expectation
value of $\vert v_s - v_c \vert$ will be larger for an interloper than
for a  true satellite. As long  as the interloper is  fainter than the
brightest  galaxy in the  group (halo)  to which  it is  assigned, its
impact on $\R$ may be small,  as it affects both the numerator and the
denominator.   However, if the  interloper is  brighter than  all true
group   members,  $\vert   \R  \vert$   will  typically   be  severely
overestimated. Another problem is related  to the fact that the 2dFGRS
and  SDSS suffer from  various incompleteness  effects. If  the actual
brightest halo galaxy is missed  (i.e., is not present in the survey),
$\R$ will be measured with  respect to a satellite galaxy, which again
will  bias $\vert  \R \vert$  high.  The  presence of  interlopers and
incompleteness effects,  therefore, tend to create  excessive wings in
the $\R$ distribution.  A comparison with the Student $t$-distribution
might  then give  the  wrong impression  that  the null-hypothesis  is
rejected.  Since  the typical occupation numbers of  haloes are small,
this   effect   can   be   very   strong,   as   we   demonstrate   in
Section~\ref{sec:res}.  To  circumvent these problems,  we compare the
$\R$-distributions obtained from groups in the 2dFGRS and SDSS against
those  obtained  from  groups  extracted  from  detailed  mock  galaxy
redshift  surveys, which  suffer from  interlopers  and incompleteness
effects to the same extent as the real data.
\begin{figure*}
\centerline{\psfig{figure=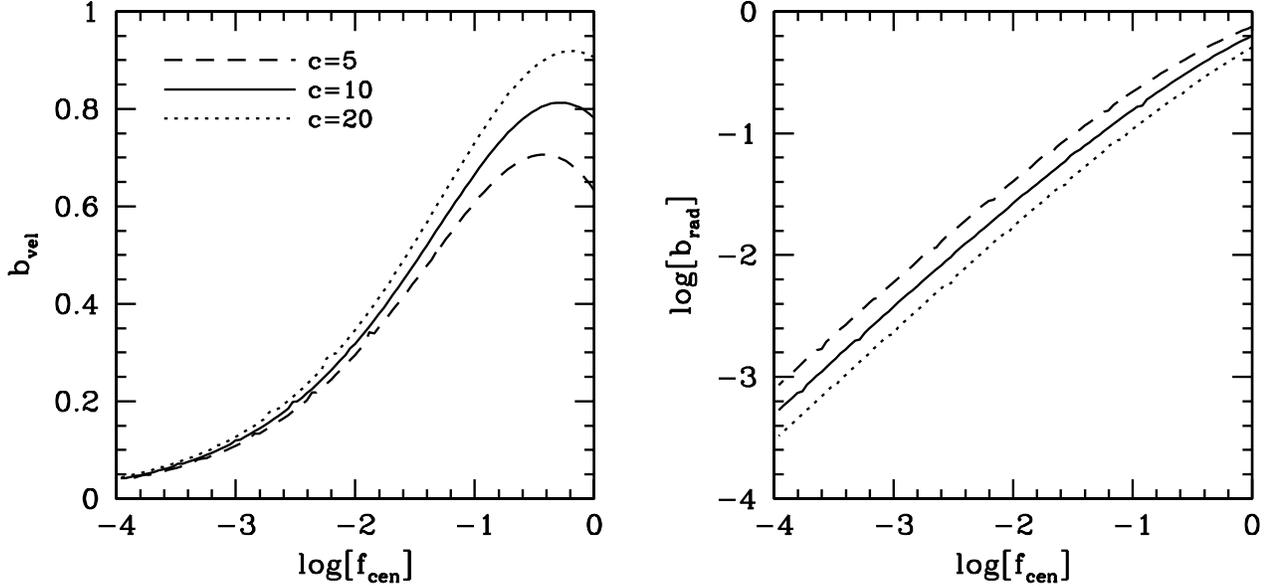,width=0.95\hdsize}}
\caption{The velocity bias (left-hand panel) and spatial bias
  (right-hand panel) of central  galaxies as function of the parameter
  $f_{\rm  cen}$,  which expresses  the  characteristic  scale of  the
  radial   distribution  of   central   galaxies  in   terms  of   the
  characteristic   scale  of   the  NFW   density   distribution  (see
  Section~\ref{sec:velbias}).  Results  are shown for  three values of
  the halo concentration parameter $c$, as indicated.}
\label{fig:cengalvbias}
\end{figure*}

\section{Application to the 2dFGRS and SDSS}
\label{sec:data}

\subsection{Group Selection}
\label{sec:groups}

The ${\cal  R}$-test described above requires a  selection of galaxies
that  belong to  the  same dark  matter  halo. In  Yang \etal  (2005a,
hereafter YMBJ) we developed a halo-based galaxy group finder, that is
optimized for this task. Here we give a brief description of this group
finder, and refer the interested reader to YMBJ for details.

The  basic idea  behind our  group finder  is similar  to that  of the
matched filter  algorithm developed by Postman  \etal (1996), although
it also makes  use of the galaxy kinematics.   The group finder starts
with an assumed mass-to-light ratio to assign a tentative mass to each
potential group, identified using the friends-of-friends (FOF) method.
This mass is used to estimate  the size and velocity dispersion of the
underlying  halo  that hosts  the  group, which  in  turn  is used  to
determine  group membership  (in redshift  space).  This  procedure is
iterated until  no further changes occur in  group memberships.  Using
detailed mock  galaxy redshift surveys,  the performance of  our group
finder has  been tested in terms  of completeness of  true members and
contamination by interlopers.   The average completeness of individual
groups  is  $\sim  90$  percent   and  with  only  $\sim  20$  percent
interlopers.    Furthermore,   the   resulting  group   catalogue   is
insensitive  to  the initial  assumption  regarding the  mass-to-light
ratios, and  is more  successful than the  conventional FOF  method in
associating galaxies according to their common dark matter haloes.

\subsection{The 2dFGRS}
\label{sec:2dfgrs}

We use  the final,  public data release  from the  2dFGRS, restricting
ourselves only to  galaxies with redshifts $0.01 \leq  z \leq 0.20$ in
the  North Galactic  Pole and  South Galactic  Pole subsamples  with a
redshift quality parameter $q \geq 3$ and a redshift completeness $c >
0.8$.   This leaves a  grand total  of $151,280$  galaxies with  a sky
coverage of $\sim 1125 \,  {\rm deg}^2$.  The typical rms redshift and
magnitude errors  are $85 \kms$ and $0.15$  mag, respectively (Colless
\etal  2001).  Absolute  magnitudes  for galaxies  in  the 2dFGRS  are
computed using the K-corrections of Madgwick \etal (2002).

Application  of the  halo-based group  finder to  this  galaxy sample,
yields  a group  catalogue consisting  of $77,708$  systems.  Detailed
information regarding the  clustering properties and galaxy occupation
statistics of these groups can be found in Yang \etal (2005a,b,c).  In
what follows  we restrict  our analyzes to  the $2502$ groups  in this
catalogue with four members or more.

\subsection{The SDSS}
\label{sec:sdss}

In  addition to  the  2dFGRS, we  also  use data  from  the SDSS.   In
particular,  we  use  the   New  York  University  Value-Added  Galaxy
Catalogue                                                    (NYU-VAGC)
\footnote{http://wassup.physics.nyu.edu/vagc/\#download}, described in
detail in  Blanton \etal  (2004).  The NYU-VAGC  is based on  the SDSS
Data Release 2 (Abazajian \etal  2004), but with an independent set of
significantly improved reductions.  From  this catalogue we select all
galaxies in the Main Galaxy  Sample, which has an extinction corrected
Petrosian magnitude  limit of $r=18$.   We prune this sample  to those
galaxies in  the redshift  range $0.01  \leq z \leq  0.20$ and  with a
redshift  completeness  $c >  0.7$.   This  leaves  a grand  total  of
$184,425$ galaxies with a sky coverage  of $\sim 1950 \, {\rm deg}^2$. 
From this  SDSS sample, we  construct a group catalogue  that contains
$102,935$ systems.  A more detailed description of this catalogue will
be presented  in Weinmann  \etal (2005, in  preparation).  As  for the
2dFGRS, we  restrict our analysis to  the groups with  four members or
more, of which there are $3473$ in our catalogue.

\subsection{Comparison of 2dFGRS with SDSS}
\label{sec:compare}

For each group in both  the 2dFGRS and SDSS catalogues described above
we compute ${\cal  R}$. Fig.~\ref{fig:comparison} plots the cumulative
distributions  of $\vert  {\cal R}  \vert$  for both  surveys. In  the
upper-left panel  we plot  the distributions using  all groups  in the
range   $50   \kms  \leq   \hat{\sigma}_s   \leq   1000  \kms$,   with
$\hat{\sigma}_s$     the    unbiased    estimator     of    $\sigma_s$
(equation~[\ref{sigest}]).   In the  other three  panels we  plot $P(<
\vert {\cal  R} \vert)$  for three sub-samples  (the values  in square
brackets  indicate the  range in  $\hat{\sigma}_s$ used,  in  $\kms$). 
Overall the agreement  between SDSS and 2dFGRS is  extremely good.  To
make the  comparison more quantitative, we  use the Kolmogorov-Smirnov
(hereafter KS) test to compute  the probability $P_{\rm KS}$ that both
$P(\vert {\cal R}  \vert)$ are drawn from the  same distribution.  The
resulting probabilities  are indicated  in each panel.   These confirm
what  can  already  be  inferred  by  eye,  namely  that  both  ${\cal
  R}$-distributions are  consistent with each other.   Given this good
agreement between both data sets, we only concentrate on the 2dFGRS in
what follows.  The main reason  for choosing this survey over the SDSS
is that  we have  accurate mocks  for the 2dFGRS  that have  been well
tested. Given  the good  agreement between 2dFGRS  and SDSS,  we argue
that any result based on the former will also hold for the latter.
\begin{figure*}
\centerline{\psfig{figure=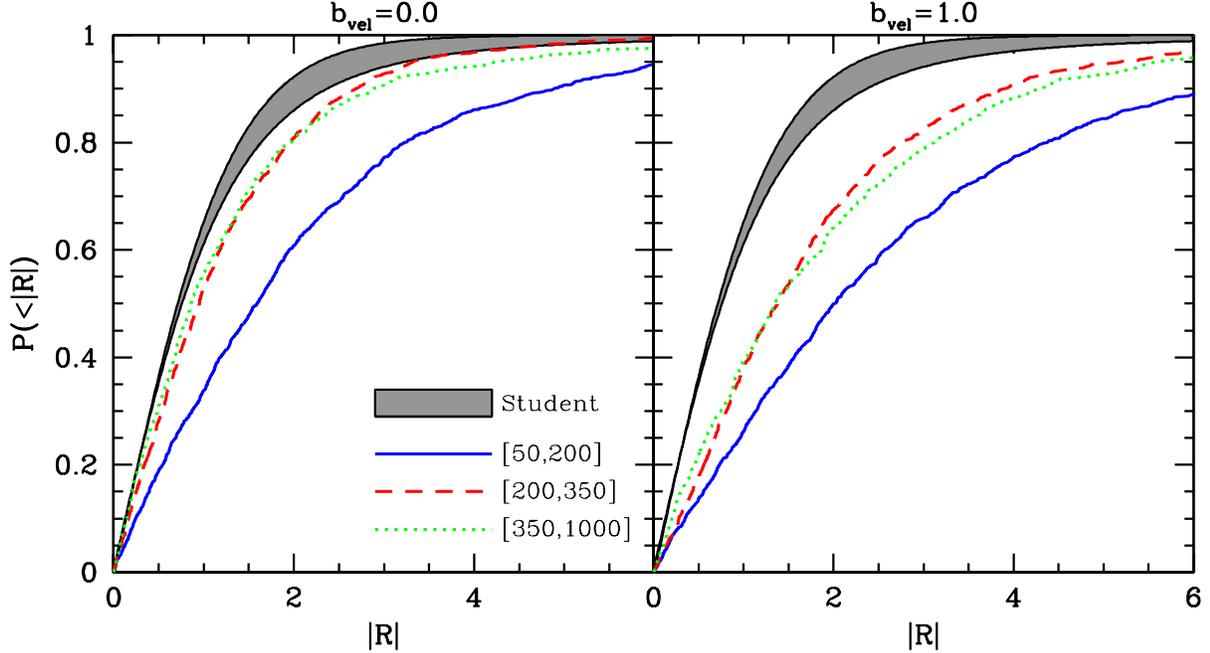,width=0.9\hdsize}}
\caption{The cumulative 
  distributions of $\vert {\cal R}\vert$ obtained from MGRSs $M_{0.0}$
  (left-hand  panel) and  $M_{1.0}$ (right-hand  panel), in  which the
  brightest halo galaxies have a velocity bias of $b_{\rm vel} =0$ and
  $b_{\rm  vel}=1$,  respectively.  Solid,  dashed  and dotted  curves
  correspond to  group samples with $50 \kms  \leq \hat{\sigma}_s \leq
  200 \kms$,  $200 \kms \leq  \hat{\sigma}_s \leq 350 \kms$,  and $350
  \kms \leq  \hat{\sigma}_s \leq  1000 \kms$, respectively.   The gray
  area indicates the area bounded  by Student distributions with 3 and
  9 degrees  of freedom.  In  the ideal case without  interlopers, the
  $P(<\vert {\cal  R}\vert)$ of $M_{0.0}$  should fall in this  range. 
  The fact that  they don't illustrates the impact  of interlopers and
  completeness effects,  and emphasizes the importance  of using MGRSs
  for a  fair comparison  with the data.   Finally, the fact  that the
  $P(<\vert   {\cal   R}\vert)$  of   $M_{0.0}$   and  $M_{1.0}$   are
  significantly  different illustrates that  the ${\cal  R}$-test does
  have  the  ability  to   constrain  the  phase-space  parameters  of
  brightest halo galaxies.}
\label{fig:studmock}
\end{figure*}

\section{Modeling velocity bias of central galaxies}
\label{sec:velbias}

The main  goal of this  paper is to  use the ${\cal  R}$ distributions
presented above  in order to  constrain the phase space  parameters of
brightest  halo galaxies.   We will  express these  in terms  of their
spatial  and velocity  bias with  respect to  the satellites.   If the
null-hypothesis  of the  CGP  is  correct, both  the  spatial and  the
velocity bias should  equal zero. In order to  model these biases, and
to incorporate them in the mock  redshift surveys that we will use for
comparison with the data, we proceed as follows.

We assume  that each dark  matter halo has  an NFW (Navarro,  Frenk \&
White  1997)  density distribution,  $\rho_{\rm  dm}(r)$, with  virial
radius   $r_{\rm  vir}$,  characteristic   scale  radius   $r_s$,  and
concentration parameter $c =  r_{\rm vir}/r_s$.  Assuming haloes to be
spherical   and  isotropic,   the   local,  one-dimensional   velocity
dispersion follows from solving the Jeans equation
\begin{equation}
\label{jeans}
\sigma^2_{\rm dm}(r) = {1 \over \rho_{\rm dm}(r)} 
\int_{r}^{\infty} \rho_{\rm dm}(r') {\partial \Psi \over \partial
  r}(r') {\rm d}r'
\end{equation}
with $\Psi(r)$ the gravitational  potential (Binney \& Tremaine 1987). 
Using that $\partial \Psi / \partial  r = G M(r)/r^2$ and defining the
virial velocity $V_{\rm vir} = \sqrt{G M / r_{\rm vir}}$ we obtain
\begin{equation}
\label{sig1dm}
\sigma^2_{\rm dm}(r) =  V^2_{\rm vir} {c \over f(c)} \, 
\left({r \over r_s}\right) \, \left(1 + {r \over r_s}\right)^2 \, 
{\cal I}(r/r_s)
\end{equation}
with  $f(x) = {\rm  ln}(1+x) -  x/(1+x)$ and
\begin{equation}
\label{cali}
{\cal I}(y) = \int_{y}^{\infty} {f(\tau) \,{\rm d}\tau 
\over \tau^3 (1+\tau)^2}\,.
\end{equation}
The halo-averaged velocity dispersion is given by
\begin{eqnarray}
\label{sigexp}
\langle \sigma_{\rm dm} \rangle_M & \equiv &
{4 \pi \over M}  \int_{0}^{r_{\rm vir}}
\rho_{\rm dm}(r) \, \sigma_{\rm dm}(r) \, r^2 \, {\rm d}r \nonumber \\
& = &V_{\rm vir} \, \sqrt{c \over f^3(c)} \,
\int_{0}^{c} {y^{3/2} \, {\cal I}^{1/2}(y) \over  
(1+y)} \, {\rm d}y
\end{eqnarray}
(cf. van den Bosch \etal 2004).
\begin{figure*}
\centerline{\psfig{figure=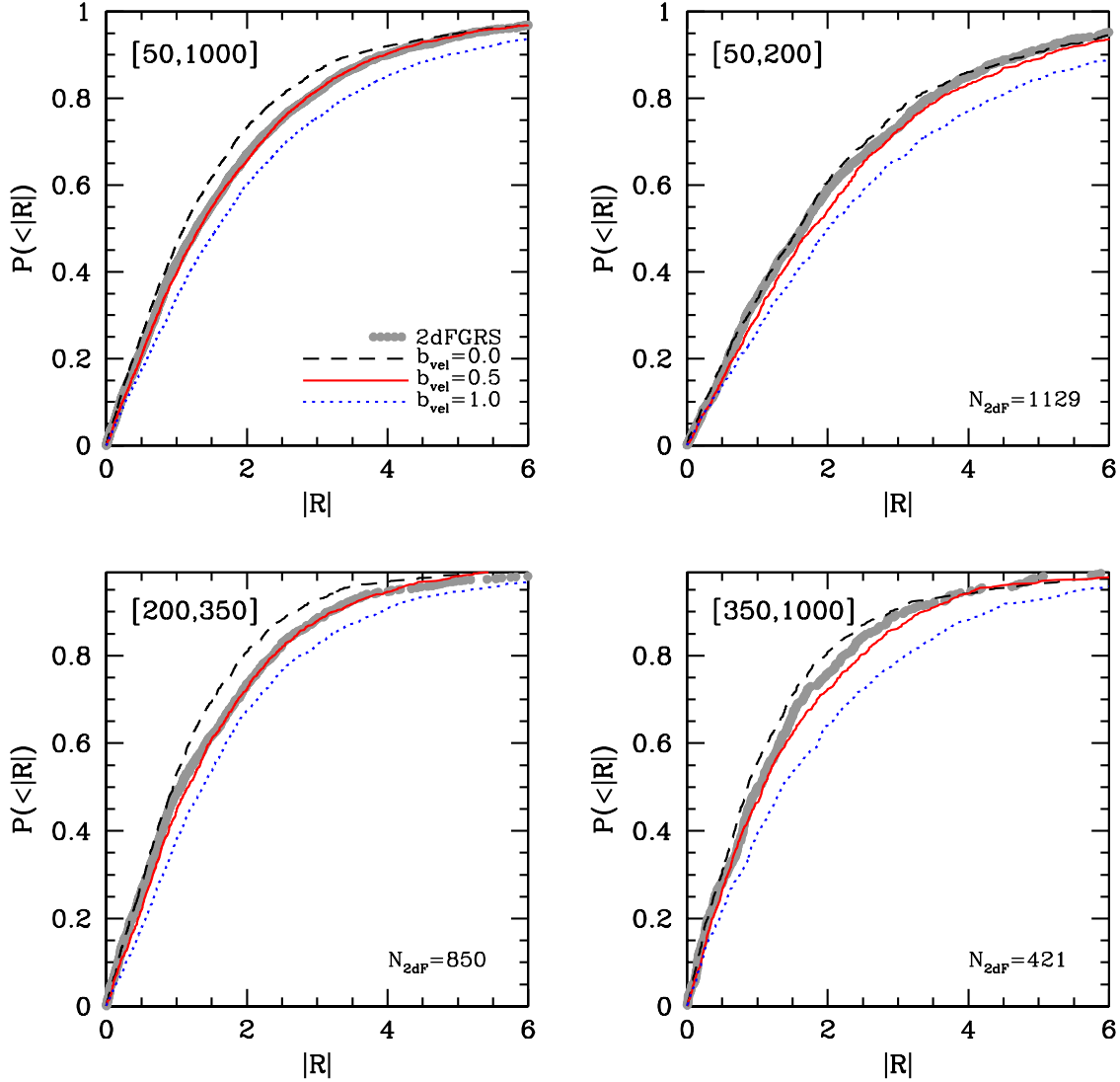,width=0.85\hdsize}}
\caption{The cumulative distributions of $\vert {\cal R} \vert$
  obtained from  the groups in  the 2dFGRS (gray dots),  compared with
  those obtained from  three of our MGRSs, which  only differ in their
  value  of $b_{\rm  vel}$,  as indicated  in  the upper  left panel.  
  Results are shown for  four intervals in $\hat{\sigma}_s$, indicated
  in square brackets  in each panel. The number of  2dF groups in each
  of the three subsamples is indicated.}
\label{fig:res}
\end{figure*}

Throughout  this paper,  we assume  that the  $N_{\rm  sat}$ satellite
galaxies in  a halo of mass  $M$ follow a  number density distribution
$n_{\rm sat}(r) = (N_{\rm sat}  / M) \rho_{\rm dm}(r)$, i.e., there is
no spatial bias between satellite  galaxies and dark matter particles. 
As shown  in van den Bosch  \etal (2005), this is  consistent with the
observed radial distribution of  satellite galaxies in the 2dFGRS.  If
we further assume that the satellites are in isotropic equilibrium, it
also follows that there is no velocity bias between the satellites and
the  dark matter,  neither globally  [i.e.  $\langle  \sigma_{\rm sat}
\rangle_M  = \langle  \sigma_{\rm dm}  \rangle_M$] nor  locally  [i.e. 
$\sigma_{\rm sat}(r) = \sigma_{\rm dm}(r)$].

When  stacking  all haloes  of  a given  mass,  we  assume that  their
brightest halo galaxies follow  a number density distribution given by
a   Hernquist   profile\footnote{The   choice  for   this   particular
  distribution is not motivated  by any physical considerations, other
  than  the fact that  it is  well behaved,  both at  $r=0$ and  at $r
  \rightarrow \infty$}:
\begin{equation}
\label{rhocen}
\rho_{\rm cen}(r) \propto {1 \over 2 \pi} {a \over r} {1 \over (r+a)^3}
\end{equation}
This  implies  a  probability  distribution  for $r$  of  the  central
galaxies of
\begin{equation}
\label{probcen}
P_{\rm cen}(r) {\rm d}r = 2 \left({r_{\rm vir} + a \over r_{\rm vir}} 
\right)^2 {a r \over (r + a)^3} {\rm d} r
\end{equation}
In order to parameterize the characteristic radius $a$ in terms of that
of the dark  matter halo, we define the  parameter $f_{\rm cen} \equiv
a/r_s$.  A brightest  halo galaxy at a halo-centric  radius $r$ has an
isotropic velocity dispersion
\begin{eqnarray}
\label{veldispcen}
\sigma^2_{\rm cen}(r) & = & {1 \over \rho_{\rm cen}(r)} 
\int_{r}^{\infty} \rho_{\rm cen}(r') {\partial \Psi \over \partial
  r}(r') {\rm d}r'\nonumber \\
 & = & V^2_{\rm vir} {c \over f(c)} \, 
\left({r \over r_s}\right) \, \left(f_{\rm cen} + {r \over
    r_s}\right)^3 {\cal J}(r/r_s)\,,
\end{eqnarray}
with
\begin{equation}
\label{calj}
{\cal J}(y) = \int_{y}^{\infty} {f(\tau) \,{\rm d}\tau 
\over \tau^3 (f_{\rm cen}+\tau)^3}\,.
\end{equation}
This implies a halo-averaged velocity dispersion of 
\begin{eqnarray}
\label{sigcenav}
\langle \sigma_{\rm cen} \rangle_M & \equiv &
{\int_{0}^{r_{\rm vir}}
\rho_{\rm cen}(r) \, \sigma_{\rm cen}(r) \, r^2 \, {\rm d}r 
\over \int_{0}^{r_{\rm vir}} \rho_{\rm cen}(r) \, r^2 \, {\rm d}r} \nonumber \\
& = &V_{\rm vir} \, \sqrt{4 c \over f(c)} \, f_{\rm cen} \,
\int_{0}^{c} {y^{3/2} \, {\cal J}^{1/2}(y) \over  
(f_{\rm cen}+y)^{3/2}} \, {\rm d}y\,,
\end{eqnarray}
which allows us to define the velocity bias of brightest halo galaxies
as  $b_{\rm vel}  \equiv \langle  \sigma_{\rm cen}  \rangle  / \langle
\sigma_{\rm dm}  \rangle = \langle \sigma_{\rm cen}  \rangle / \langle
\sigma_{\rm sat} \rangle$. In addition to the velocity bias, we define
the spatial bias as $b_{\rm  rad} \equiv \langle r_{\rm cen} \rangle /
\langle r_{\rm  dm} \rangle  = \langle r_{\rm  cen} \rangle  / \langle
r_{\rm  sat} \rangle$,  where  the expectation  value  for the  radius
follows from
\begin{equation}
\label{rexpect}
\langle r \rangle = {\int_{0}^{r_{\rm vir}} \rho(r) r^3 {\rm d}r 
\over \int_{0}^{r_{\rm vir}} \rho(r) r^2 {\rm d}r}
\end{equation}
For an  NFW density distribution with concentration  $c$, this reduces
to
\begin{eqnarray}
\label{rnfwexp}
\langle r_{\rm dm} \rangle = \left[{(2+c)/(1+c) - (2/c) {\rm ln}(1+c) 
\over {\rm ln}(1+c) - c/(1+c)}\right] \, r_{\rm vir}
\end{eqnarray}
or  $\langle  r_{\rm dm}  \rangle  = 0.41  r_{\rm  vir}$  for $c=10$.  

Fig.~\ref{fig:cengalvbias} plots  $b_{\rm vel}$ (left-hand  panel) and
$b_{\rm  rad}$ (right-hand  panel) as  function of  $f_{\rm  cen}$ for
three values  of the  halo concentration parameter  $c$. In  the limit
$f_{\rm  cen}  \rightarrow 0$,  the  probability distribution  $P_{\rm
  cen}(r)$  becomes a  Dirac delta  function.  This  implies  that the
central galaxy is sitting still at  the center of the dark matter halo
(i.e., the null-hypothesis of the  CGP), so that $b_{\rm vel} = b_{\rm
  rad} =  0$.  Increasing $f_{\rm  cen}$ increases the  probability to
find  the brightest halo  galaxy at  larger halo-centric  radii, which
corresponds to  a larger velocity  bias.  Note, however,  that $b_{\rm
  vel}$  never  approaches  unity,  which  is due  to  the  fact  that
$\rho_{\rm cen}(r)$  can not be  made to match $\rho_{\rm  dm}(r)$ for
any value of $f_{\rm cen}$.   Typically $b_{\rm vel} \gg b_{\rm rad}$,
which  is a  reflection  of the  `depth'  of the  NFW potential.   For
example,  for  a   velocity  bias  of  $b_{\rm  vel}   =  0.5$  (i.e.,
corresponding to a specific kinetic energy that is one quarter of that
of  the satellites)  the  radial  bias is  $b_{\rm  rad} \simeq  0.07$
(assuming  $c=10$). Combining  this with  ~(\ref{rnfwexp})  implies an
expectation value for  the offset of the central  galaxy from the dark
matter distribution of $\langle r_{\rm cen} \rangle \simeq 0.03 r_{\rm
  vir}$.  For  a Milky-Way  sized system this  corresponds to  $\sim 5
\kpc$, comparable  to the characteristic radius (scale  length) of the
galaxy itself.

In what  follows, we construct a  set of mock  galaxy redshift surveys
(hereafter MGRSs)  for different values of $b_{\rm  vel}$, and compare
the  ${\cal R}$-distributions  of their  groups against  those  of the
2dFGRS, which are statistically identical  to those of the SDSS, in an
attempt to constrain $b_{\rm vel}$.
\begin{table*}
\caption{Comparison between MGRSs and 2dFGRS}
\begin{tabular}{cccccccc}
   \hline
MGRS & $b_{\rm vel}$ & $b_{\rm rad}$ & $f_{\rm cen}$ &
$P_{\rm KS}[50,1000]$ & $P_{\rm KS}[50,200]$ & 
$P_{\rm KS}[200,350]$ & $P_{\rm KS}[350,1000]$ \\
 (1) & (2) & (3) & (4) & (5) & (6) & (7) & (8) \\ 
\hline\hline
${\rm M}_{0.0}$ & $0.0$ & $0.0$               & $0.0$               & $1.5\times 10^{-6}$ & $2.6\times 10^{-1}$ & $5.1\times 10^{-4}$ & $2.9\times 10^{-2}$ \\
${\rm M}_{0.1}$ & $0.1$ & $2.8\times 10^{-3}$ & $7.3\times 10^{-4}$ & $6.0\times 10^{-4}$ & $8.7\times 10^{-1}$ & $2.3\times 10^{-1}$ & $7.4\times 10^{-2}$ \\
${\rm M}_{0.2}$ & $0.2$ & $1.0\times 10^{-2}$ & $3.3\times 10^{-3}$ & $3.7\times 10^{-3}$ & $4.9\times 10^{-1}$ & $6.2\times 10^{-1}$ & $8.7\times 10^{-2}$ \\
${\rm M}_{0.3}$ & $0.3$ & $2.4\times 10^{-2}$ & $8.5\times 10^{-3}$ & $4.3\times 10^{-3}$ & $4.3\times 10^{-1}$ & $2.4\times 10^{-1}$ & $1.6\times 10^{-1}$ \\
${\rm M}_{0.4}$ & $0.4$ & $4.2\times 10^{-2}$ & $1.8\times 10^{-2}$ & $1.8\times 10^{-1}$ & $8.2\times 10^{-2}$ & $1.3\times 10^{-1}$ & $3.5\times 10^{-1}$ \\
${\rm M}_{0.5}$ & $0.5$ & $7.2\times 10^{-2}$ & $3.5\times 10^{-2}$ & $2.8\times 10^{-1}$ & $7.5\times 10^{-2}$ & $1.0\times 10^{-1}$ & $3.8\times 10^{-1}$ \\
${\rm M}_{0.6}$ & $0.6$ & $1.1\times 10^{-1}$ & $6.6\times 10^{-2}$ & $8.5\times 10^{-3}$ & $4.0\times 10^{-2}$ & $1.1\times 10^{-3}$ & $1.2\times 10^{-1}$ \\
${\rm M}_{0.7}$ & $0.7$ & $1.8\times 10^{-1}$ & $1.3\times 10^{-1}$ & $3.3\times 10^{-3}$ & $3.3\times 10^{-3}$ & $1.2\times 10^{-2}$ & $7.0\times 10^{-2}$ \\
${\rm M}_{0.8}$ & $0.8$ & $3.4\times 10^{-1}$ & $3.3\times 10^{-1}$ & $1.0\times 10^{-5}$ & $2.4\times 10^{-5}$ & $1.1\times 10^{-3}$ & $7.4\times 10^{-3}$ \\
${\rm M}_{1.0}$ & $1.0$ & $1.0$               & $--$                & $1.6\times 10^{-9}$ & $1.9\times 10^{-5}$ & $1.3\times 10^{-6}$ & $5.3\times 10^{-6}$ \\
\hline
\end{tabular}
\medskip

\begin{minipage}{\hdsize}
  The MGRSs used for comparison with the 2dFGRS.  Column~(1) lists the
  ID of  the MGRS. Columns (2),  (3), and (4) list  the velocity bias,
  spatial  bias,  and  value   of  $f_{\rm  cen}$,  respectively  (see
  Section~\ref{sec:velbias}   for   definitions).   Finally,   columns
  (5)--(8) list   the   KS  probabilities   $P_{\rm   KS}$  that   the
  distributions of ${\cal R}$ extracted from these MGRS are consistent
  with  those   of  the  2dFGRS   for  four  different   intervals  in
  $\hat{\sigma}_s$,  indicated by  the values  in square  brackets (in
  $\kms$).
\end{minipage}

\end{table*}

\section{Mock Galaxy Redshift Surveys}
\label{sec:mock}

We construct MGRSs  by populating dark matter haloes  with galaxies of
different  luminosities. The  distribution  of dark  matter haloes  is
obtained from a  set of large $N$-body simulations  (dark matter only)
for  a $\Lambda$CDM  `concordance'  cosmology with  $\Omega_m =  0.3$,
$\Omega_{\Lambda}=0.7$, $h=0.7$ and  $\sigma_8=0.9$.  In this paper we
use two simulations with $N=512^3$ particles each, which are described
in more detail  in Jing \& Suto (2002).  The simulations have periodic
boundary  conditions and box  sizes of  $L_{\rm box}=100  h^{-1} \Mpc$
(hereafter  $L_{100}$) and  $L_{\rm box}=300  h^{-1}  \Mpc$ (hereafter
$L_{300}$). We  follow Yang \etal  (2004) and replicate  the $L_{300}$
box on a $4  \times 4 \times 4$ grid.  The central  $2 \times 2 \times
2$ boxes, are  replaced by a stack of $6 \times  6 \times 6$ $L_{100}$
boxes, and the  virtual observer is placed at  the center (see Fig.~11
in   Yang   \etal   2004).    This   stacking   geometry   circumvents
incompleteness problems  in the mock  survey due to  insufficient mass
resolution of  the $L_{300}$ simulations,  and allows us to  reach the
desired depth of $z_{\rm max}=0.20$ in all directions.
 
Dark  matter haloes are  identified using  the standard  FOF algorithm
with  a  linking  length   of  $0.2$  times  the  mean  inter-particle
separation.  Unbound haloes and haloes with less than 10 particles are
removed from the sample.  In Yang  \etal (2004) we have shown that the
resulting  halo mass  functions are  in excellent  agreement  with the
analytical halo mass function of Sheth, Mo \& Tormen (2001).

\subsection{Populating Haloes with Galaxies}
\label{sec:hon}

In order to populate the dark matter haloes with galaxies of different
luminosities,  we use the  conditional luminosity  function (hereafter
CLF), $\Phi(L \vert M)$, which gives the average number of galaxies of
luminosity $L$ that resides in a  halo of mass $M$. As demonstrated in
Yang, Mo \& van den Bosch (2003) and van den Bosch, Yang \& Mo (2003),
the CLF is  well constrained by the galaxy  luminosity function and by
the galaxy-galaxy  correlation lengths  as function of  luminosity. In
the MGRSs used  here we use the CLF  with ID \# 6 given  in Table~1 of
van den  Bosch \etal (2005). We  have tested that none  of our results
depend significantly on this particular choice for the CLF.

Because of the  mass resolution of the simulations  and because of the
completeness limit of the 2dFGRS, we adopt a minimum galaxy luminosity
of  $L_{\rm min} =  10^{7} h^{-2}  \Lsun$.  The  {\it mean}  number of
galaxies with $L \geq L_{\rm min}$  that resides in a halo of mass $M$
is given by
\begin{equation}
\label{totN}
\langle N \rangle_M = \int_{L_{\rm min}}^{\infty} \Phi(L \vert M) \, {\rm d}L
\end{equation}
In  order  to Monte-Carlo  sample  occupation  numbers for  individual
haloes, one requires the  full probability distribution $P(N \vert M)$
(with $N$ an  integer) of which $\langle N \rangle_M$  gives the mean. 
We differentiate between satellite  galaxies and central galaxies. The
total number  of galaxies per  halo is the  sum of $N_{\rm  cen}$, the
number of  central galaxies which is  either one or  zero, and $N_{\rm
  sat}$, the (unlimited) number of satellite galaxies.  We assume that
$N_{\rm sat}$ follows a  Poisson distribution and require that $N_{\rm
  sat}=0$ whenever $N_{\rm  cen}=0$.  The halo occupation distribution
is thus  specified as  follows: if $\langle  N \rangle_M \leq  1$ then
$N_{\rm sat} =  0$ and $N_{\rm cen}$ is  either zero (with probability
$P = 1 - \langle N \rangle_M$) or one (with probability $P = \langle N
\rangle_M$).  If  $\langle N \rangle_M  > 1$ then $N_{\rm  cen}=1$ and
$N_{\rm  sat}$ is drawn  from a  Poisson distribution  with a  mean of
$\langle N \rangle_M - 1$.

We follow Yang  \etal (2004) and draw the  luminosity of the brightest
galaxy in each halo from  $\Phi(L \vert M)$ using the restriction that
$L > L_1$ with $L_1$ defined by
\begin{equation}
\label{Lons}
\int_{L_1}^\infty \Phi(L\vert M) dL = 1\,.
\end{equation}
The  luminosities  of  the  satellite  galaxies are  also  drawn  from
$\Phi(L\vert M)$, but with the restriction $L_{\rm min} < L < L_1$.

Next we assign all galaxies a position and velocity within their halo,
using  the  number  density  distributions  and  (isotropic)  velocity
dispersion profiles given in Section~\ref{sec:velbias}. Note that this
implicitly  assumes  that  all   haloes,  as  well  as  their  galaxy
populations, are relaxed. Halo concentrations as function of halo mass
are computed  using the  relation given by  Eke, Navarro  \& Steinmetz
(2001).

\subsection{Creating Mock Surveys}
\label{sec:stack}

The  2dFGRS  uses  a  multifibre  spectrograph to  obtain  redshifts.  
However, because of the physical size of the fibers, when two galaxies
are closer than $\sim 30$ arcsec in projection only one of them can be
targeted.   Furthermore, due  to  clustering, some  areas  on the  sky
contain  more  galaxies within  a  single  two-degree  field than  the
available number of fibers.   By using a sophisticated tiling strategy
these  problems  are  largely  overcome,  yielding  a  fairly  uniform
sampling rate.  Nevertheless, some spatial non-uniformities remain. In
addition, fainter  galaxies yield noisier spectra,  and therefore less
accurate redshifts.   All these effects combined result  in a redshift
completeness which depends on both position on the sky and on apparent
magnitude.   The 2dFGRS  team has  constructed maps  that parameterize
this  position  and magnitude  dependent  completeness (Colless  \etal
2001; Norberg \etal 2002), and  which facilitate a simulation of these
effects  in our  MGRSs. However,  as  it turns  out, the  completeness
depends also on the  angular separation, $\theta$, between galaxy {\it
  pairs} (see Hawkins \etal 2003).  This is largely due to the problem
of fiber  collisions, which has  not been completely corrected  for by
the tiling  strategy.  Finally, Norberg  \etal (2002) have  shown that
the {\it parent}  catalogue of the 2dFGRS, the  APM catalogue, is only
91\%  complete.   As  shown  in  van  den  Bosch  \etal  (2005),  this
incompleteness is,  at least partially,  due to image blending  in the
APM catalogue (see also Cole \etal 2001). Based on this information we
mimic the various observational  selection and completeness effects in
the 2dFGRS using the following steps:

\begin{enumerate}
  
\item We  define a $(\alpha,\delta)$-coordinate frame  with respect to
  the virtual observer at the center of the stack of simulation boxes,
  and remove all galaxies that are not located in the areas equivalent
  to the NGP and SGP regions of the 2dFGRS.
  
\item For each  galaxy we compute the apparent  magnitude according to
  its luminosity  and distance, to  which we add  a rms error  of 0.15
  mag.  Since galaxies in the 2dFGRS were pruned by apparent magnitude
  {\it before} a  K-correction was applied, we proceed  as follows: We
  first apply a negative  K-correction, then select galaxies according
  to  the  position-dependent  magnitude  limit  (obtained  using  the
  apparent  magnitude limit masks  provided by  the 2dFGRS  team), and
  finally   K-correct  the   magnitudes  back   to   their  rest-frame
  $b_J$-band.   Throughout  we  use the  type-dependent  K-corrections
  given in Madgwick \etal (2002).
  
\item For each galaxy we compute the redshift as `seen' by the virtual
  observer.   We take  the observational  velocity  uncertainties into
  account  by   adding  a  random  velocity  drawn   from  a  Gaussian
  distribution with dispersion $85\kms$.
  
\item To  take  account  of  the  position-  and  magnitude-dependent
  completeness of the 2dFGRS, we randomly sample each galaxy using the
  completeness masks provided by the 2dFGRS team.
    
\item To  take account of the  fiber-collision induced incompleteness,
  we compute the angular separations $\theta$ between all galaxy pairs
  and  remove galaxies based  on a  probability $p(\theta)$,  which we
  tune (by trial  and error) so that we  reproduce the pair-separation
  incompleteness quantified by Hawkins \etal (2003).
  
\item To take  account of the incompleteness in  the APM catalogue due
  to image blending we model the characteristic size of a galaxy as
\begin{equation}
\label{size}
R_{\rm gal} = 15 h^{-1} \kpc \, 
  \left( {L \over 10^{10} h^{-2} \Lsun} \right)^{1/3}
\end{equation}
and define  the critical projection  angle $\theta_{\rm max}  = R_{\rm
  gal} / D_A$, with $D_A$ the angular diameter distance of the galaxy.
We then remove the faintest galaxy  from all pairs for which $\theta <
\theta_{\rm max}$.

\item Finally, we remove a  number of galaxies completely at random to
  bring  the  total  fraction  of removed  galaxies,  including  those
  removed under (v) and (vi), to 9 percent.

\end{enumerate}

As shown in van den Bosch \etal (2005), this procedure results in mock
2dFGRS   catalogues   that   accurately   mimic   all   the   various
incompleteness effects,  allowing for a  direct, one-to-one comparison
with the true 2dFGRS.
\begin{figure*}
\centerline{\psfig{figure=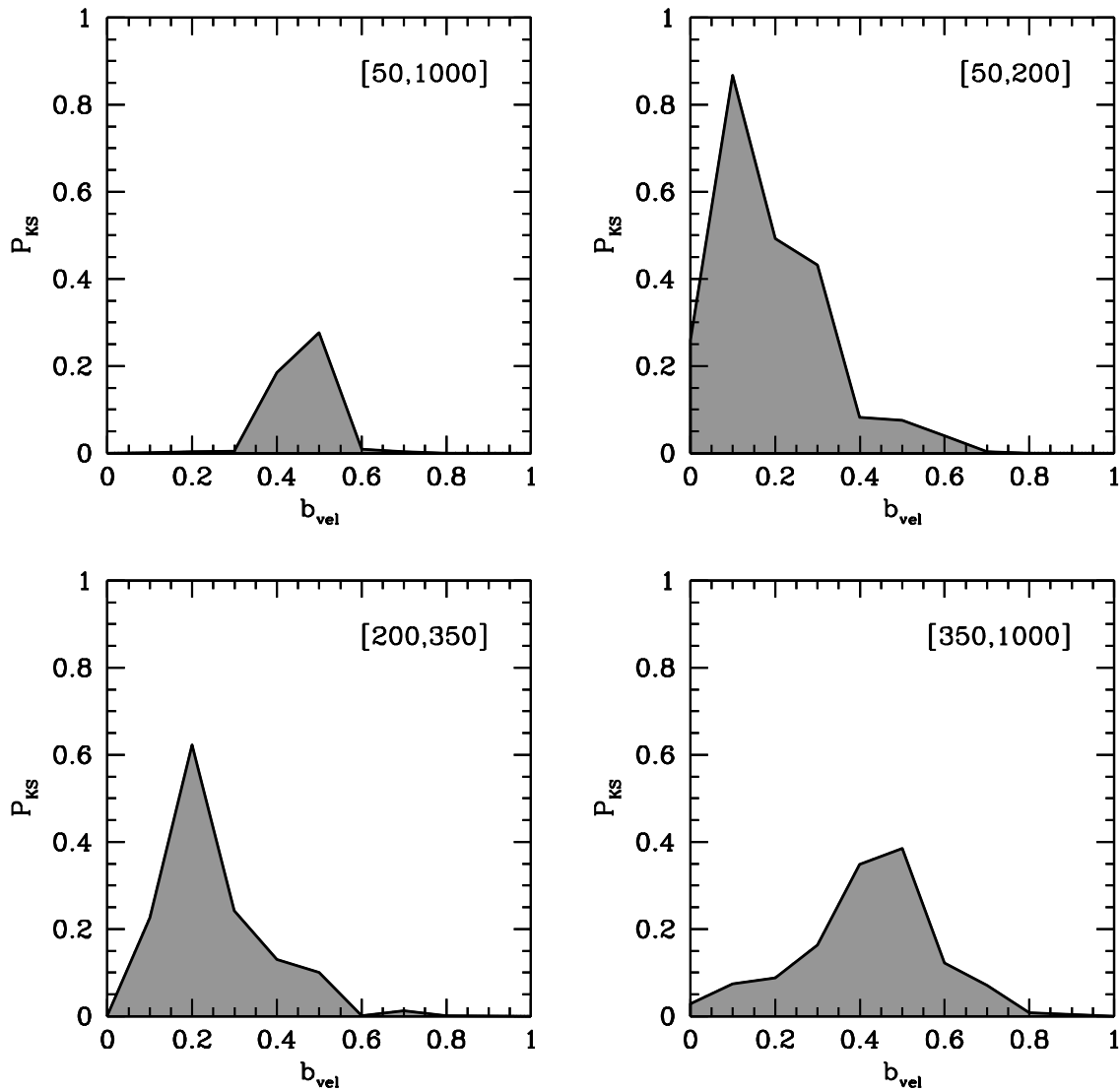,width=0.85\hdsize}}
\caption{The KS-probability that the ${\cal R}$ distribution obtained
  from the  2dFGRS groups  is consistent with  that obtained  from our
  MGRSs, as  function of  $b_{\rm vel}$.  Results  are shown  for four
  $\hat{\sigma}_s$-intervals,  indicated in  square  brackets in  each
  panel.}
\label{fig:ks}
\end{figure*}

\section{Results}
\label{sec:res}

Using the method outlined above, we  construct a set of ten MGRSs that
only differ in the value  of $b_{\rm vel}$.  Table~1 lists these mocks
together with their corresponding  values of $b_{\rm rad}$ and $f_{\rm
  cen}$, computed  for a halo  with a concentration parameter  $c=10$. 
In  the  case  of  ${\rm  M}_{1.0}$, we  deviated  somewhat  from  the
procedure described  in Section~\ref{sec:velbias}. Rather  than giving
the  central galaxies a  probability distribution  (\ref{probcen}), we
simply  treat the  brightest  galaxy  as a  satellite  galaxy so  that
$b_{\rm vel}  = 1.0$.   Note that  in this case  $f_{\rm cen}$  is not
defined.  For each of our ten MGRSs, we construct a group catalogue as
described in Section~\ref{sec:groups},  using those mock galaxies that
are  in  the  redshift range  $0.01  \leq  z  \leq  0.20$ and  with  a
completeness $c  > 0.8$ (this mimics  our selection from  the 2dFGRS). 
In what follows  we restrict our analysis to groups  with four or more
members and with $50 \kms \leq \hat{\sigma}_s \leq 1000 \kms$.

To illustrate  the importance of  using MGRSs, the left-hand  panel of
Fig.~\ref{fig:studmock} shows  the cumulative distributions  of $\vert
{\cal R} \vert$ obtained from  the groups in ${\rm M}_{0.0}$ for which
$b_{\rm vel}  = 0.0$.  In this  mock all brightest  halo galaxies have
been  located at  rest  at the  center  of the  halo.   The gray  area
indicates the area  bounded by Student $t$-distributions with  3 and 9
degrees of freedom (corresponding to systems with 4 and 10 satellites,
respectively, which  spans the range  covered by the vast  majority of
our groups).  In principle,  since this MGRS obeys the null-hypothesis
of the CGP, the resulting $P(<\vert{\cal R}\vert)$ should fall in this
range. Clearly  it doesn't,  especially not for  groups with  $50 \kms
\leq \hat{\sigma}_s  \leq 200  \kms$ (solid line).   This owes  to the
completeness effects  in the  survey, and to  the fact that  our group
finder is  not perfect and  (unavoidably) selects interlopers.   As we
discussed  in  Section~\ref{sec:sign},  these  effects  systematically
broaden  $P({\cal  R})$ (see  discussion  in Section~\ref{sec:sign}).  
Since the  impact of one  or two interlopers  is much stronger  in low
mass groups,  which have  fewer members, the  $P({\cal R})$  of groups
with  low $\hat{\sigma}_s$  deviates more  from the  predicted Student
$t$-distribution  than  that of  more  massive  groups.  This  clearly
demonstrates  that  one needs  to  take  interlopers and  completeness
effects into  account, in a  statistical sense, when  interpreting the
distribution of ${\cal  R}$ obtained from the 2dFGRS.   The MGRSs used
here are ideally suited for this task.

The right-hand panel of Fig.~\ref{fig:studmock} shows the same results
as in the left-hand panel, but  now based on ${\rm M}_{1.0}$ for which
$b_{\rm  vel}=1.0$.  Clearly,  for  this MGRS  the  $P({\cal R})$  are
significantly  broader than  for ${\rm  M}_{0.0}$.   This demonstrates
that,  despite  the   interloper/completeness  problem,  the  detailed
distributions of ${\cal R}$  obtained from group catalogues do contain
useful information that we can use to constrain $b_{\rm vel}$.

In Fig.~\ref{fig:res} we compare  $P(< \vert {\cal R} \vert)$ obtained
from our  2dFGRS group catalogue  (gray dots), to those  obtained from
three MGRSs with different values  of $b_{\rm vel}$, as indicated. The
upper-left panel plots the results  using all groups in the full range
of $\hat{\sigma}_s$ considered.  Clearly, the MGRS with $b_{\rm vel} =
0.0$  (i.e.  the  one that  fulfills the  null-hypothesis of  the CGP)
predicts a narrower  distribution of $\vert {\cal R}  \vert$ than that
found for the  2dFGRS.  Although a model in  which there is absolutely
no segregation  of the  brightest galaxy, i.e.,  $b_{\rm vel}  = 1.0$,
predicts a  distribution that is  clearly too broad,  the intermediate
case,  with $b_{\rm vel}  = 0.5$  matches the  2dFGRS results  nicely. 
Similar, though  somewhat more noisy  results (because of  the smaller
number of groups involved) are obtained for the three separate bins of
$\hat{\sigma}_s$ shown in the other three panels.

Fig.~\ref{fig:ks}  plots the  KS probability,  $P_{\rm KS}$,  that the
$P(\vert {\cal  R} \vert)$ of the  2dFGRS and the MGRS  are drawn from
the  same  distribution,  as  function  of  $b_{\rm  vel}$  (see  also
Table~1).  This analysis shows that the brightest halo galaxies in the
2dFGRS (and thus also the  SDSS) have a non-zero velocity with respect
to the  coordinate frame in which  the mean satellite motion  is zero. 
When using all  groups in the range $50  \kms \leq \hat{\sigma}_s \leq
1000  \kms$, the  data  is most  consistent  with a  velocity bias  of
$b_{\rm  vel} \simeq 0.5$,  while the  null hypothesis  of the  CGP is
rejected  at a  high level  of confidence  ($P_{\rm KS}  =  1.5 \times
10^{-6}$,    cf.      Table~1).     When    analyzing     the    three
$\hat{\sigma}_s$-subsamples, their  is hint that the  velocity bias of
brightest halo galaxies is more pronounced in more massive haloes.  We
caution, however, that these results are more noisy due to the smaller
number statistics.
\begin{figure*}
\centerline{\psfig{figure=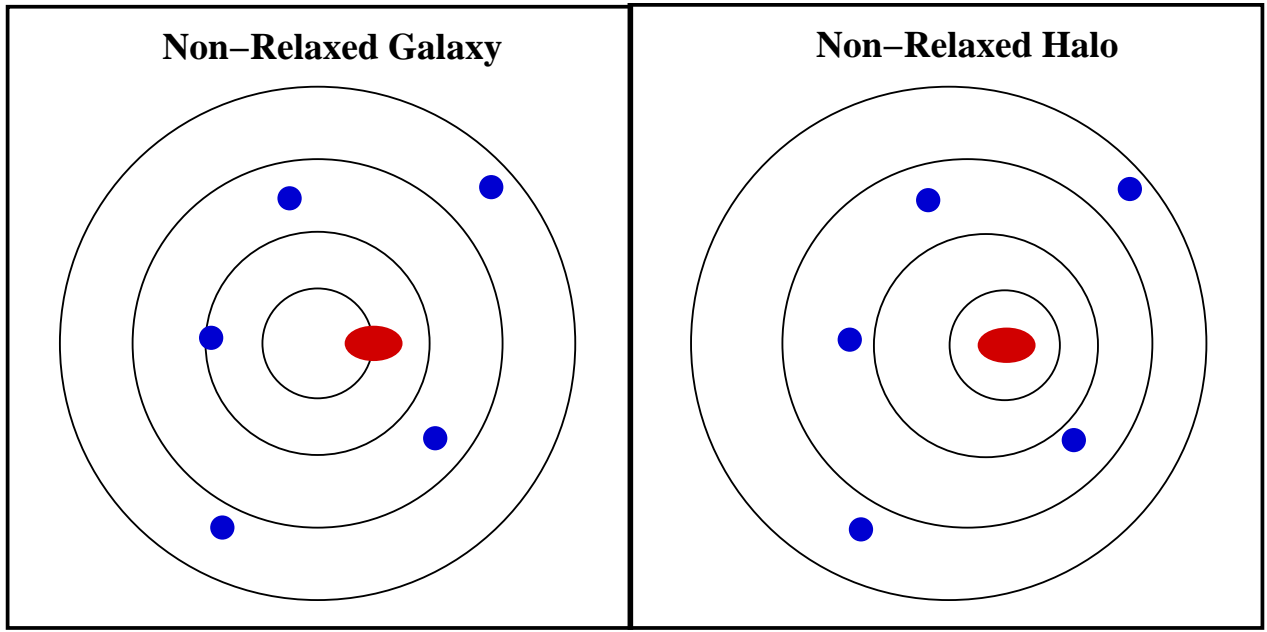,width=0.95\hdsize}}
\caption{An illustration of the two different configurations that are
  both consistent with our  inferred offset between the brightest halo
  galaxy and the satellite galaxies. Contours depict equipotentials of
  the dark matter haloes,  while filled ellipses and circles represent
  brightest halo galaxies and satellite galaxies, respectively. In the
  Non-Relaxed  Galaxy  (NRG)   scenario,  the  brightest  halo  galaxy
  oscillates in  a fully relaxed  halo. In the Non-Relaxed  Halo (NRH)
  scenario, on the  other hand, the central galaxy  coincides with the
  minimum  of  the  halo  potential,  but  the  centers  of  different
  equipotential surfaces are  offset from each other.  See  text for a
  detailed discussion.}
\label{fig:scenarios}
\end{figure*}

\section{Discussion}
\label{sec:disc}

Our finding that brightest halo  galaxies have, on average, a specific
kinetic energy that is $\sim 25$ percent of that of satellite galaxies
has two possible interpretations. First of all, the central galaxy may
not be at rest with respect  to the {\it virialized} dark matter halo. 
This scenario, which  we hereafter refer to as  the Non-Relaxed Galaxy
(NRG)   scenario,   is  illustrated   in   the   left-hand  panel   of
Fig.~\ref{fig:scenarios}.   The right-hand  panel  depicts the  second
possible scenario;  that of a  Non-Relaxed Halo (NRH).  In  this case,
the brightest halo galaxy is located  at rest with respect to the {\it
  minimum}  of the  dark matter  potential, but  the dark  matter mass
distribution is not relaxed and  reveals a clear $m=1$ mode (i.e., the
potential  minimum does not  coincide with  the barycenter).   In both
scenarios, the brightest  halo galaxy has a net  velocity with respect
to the coordinate frame defined by  the mean motion of the satellites. 
Note that, although our MGRSs are  based on the NRG scenario, to first
order it also mimics the  NRH scenario, so that our comparison between
MGRS and 2dFGRS applies to both cases.

Of the two scenarios illustrated in Fig.~\ref{fig:scenarios}, the most
likely one  is the NRH scenario.   It seems to fit  naturally within a
hierarchical picture of structure  formation, where haloes continue to
grow in mass  by accretion and merging. It is also  in accord with our
finding that  $b_{\rm vel}$  is larger in  more massive  haloes, which
form  later and are  thus expected  to be  less relaxed.   A potential
problem  for  this  scenario,  however,   is  the  fact  that  in  the
$\Lambda$CDM  concordance  cosmology  the  growth rate  of  structures
should drop fairly rapidly at the current epoch. Numerical simulations
are  ideally suited  to investigate  whether a  value of  $b_{\rm vel}
\simeq  0.5$  is a  natural  outcome  of  structure formation  in  the
$\Lambda$CDM  concordance cosmology  or  not. Using  a combination  of
numerical simulations  and semi-analytical models  of galaxy formation
in  a $\Lambda$CDM  cosmology, Diaferio  \etal (1999)  found  that the
central  galaxy has  an  average  velocity with  respect  to the  halo
barycenter  of $\sim  80  \kms$.  With  a  median halo  mass of  $\sim
10^{13} h^{-1} \Msun$, this corresponds to an average velocity bias of
$b_{\rm vel} \simeq  0.35$, in reasonable agreement with  our results. 
Yoshikawa,   Jing  \&   B\"orner  (2003)   used   a  smoothed-particle
hydrodynamics (SPH)  simulation of galaxy formation  in a $\Lambda$CDM
universe, and  found that the average velocity  difference between the
most massive halo  galaxy and the mass center of  the dark matter halo
is $\langle  v_c - v_{\rm  dm} \rangle \simeq 0.5  \langle \sigma_{\rm
  dm}  \rangle$ (see  their Fig.~10)\footnote{We  assumed  isotropy to
  convert the three-dimensional velocity dispersion of the dark matter
  particles quoted  in Yoshikawa  \etal (2003) to  the one-dimensional
  velocity dispersion $\langle \sigma_{\rm  dm} \rangle$.}. This is in
excellent  agreement with  our best-fit  value of  $b_{\rm  vel}=0.5$. 
These two simulations  suggest that our results are  in perfect accord
with  the  $\Lambda$CDM concordance  cosmology.   Note, however,  that
Yoshikawa \etal  (2003) did not  investigate whether the  most massive
halo galaxy  has a net  velocity with respect  to the most  bound halo
particle,  while  in  Diaferio  \etal  (1999) the  central  galaxy  is
associated  with  the  most  bound  halo particles  by  construction.  
Therefore, we not use either  of these results to discriminate between
the NRG and NRH scenarios.

The  NRG  scenario  appears  unlikely  at first  sight,  as  dynamical
friction  against  the highly  concentrated  dark  matter halo  should
quickly damp any oscillatory motion. On the other hand, if dark matter
haloes are cored, rather than cusped, the oscillations may persist for
a  much  longer  time  (cf.   Bontekoe  1988).   This  possibility  is
interesting  in light  of various  independent claims  for  cored dark
matter  haloes, based  on rotation  curves  of dwarf  and low  surface
brightness  galaxies  (e.g.,  Moore  1994;  Flores  \&  Primack  1994;
Borriello \& Salucci 2001; de Blok  \etal 2001; de Blok \& Bosma 2002,
but see also van den Bosch  \etal 1999; van den Bosch \& Swaters 2001;
Dutton \etal 2005), on the  observed pattern speeds of barred galaxies
(Debattista  \& Sellwood  1998, 2000),  and  on the  longevity of  the
lopsidedness of disk galaxies (Levine \& Sparke 1998). A more in-depth
study of the damping rate of these kind of oscillations in dark matter
haloes, both cusped and cored, could shed more light on these issues.

Independent of  which of  the aforementioned scenarios  is responsible
for the non-zero velocity bias  of the brightest halo galaxies, it has
important implications  for various  areas in astrophysics.   First of
all, it has an important impact  on the use of satellite kinematics to
infer  halo masses.   Since  the number  of  detectable satellites  in
individual systems  is generally small, one typically  stacks the data
on many host-satellite pairs  to obtain {\it statistical} estimates of
halo masses (Erickson, Gottesman  \& Hunter 1987; Zaritsky \etal 1993,
1997; Zaritsky  \& White 1994;  McKay \etal 2002; Brainerd  \& Specian
2003; Prada \etal  2003; van den Bosch \etal 2004).   The halo mass is
typically  derived  from the  dispersion,  $\sigma_{\rm  cs}$, of  the
distribution  of the  velocity difference  between host  and satellite
galaxies.  This  derivation rests  on the (standard)  assumptions that
the host galaxies (i.e., the brightest halo galaxies) are at rest with
respect to  the center of  a {\it relaxed}  dark matter halo.   If the
satellite galaxies have the  same kinematics as dark matter particles,
then $\sigma_{\rm cs} =  \sigma_{\rm dm} \propto M^{1/3}$. However, in
the  case of  the NRH  scenario, one  simply can  not  use (satellite)
kinematics to infer reliable halo masses, as the crucial assumption of
a virialized system is not correct.   In the case of the NRG scenario,
on the other hand, the system  is relaxed but, because of the non-zero
velocity bias, we have that  $\sigma_{\rm cs} = \sqrt{1 + b_{\rm vel}}
\sigma_{\rm dm}$.  If  one were not to correct  for $b_{\rm vel}$, the
inferred  halo  mass will  be  overestimated  by  a factor  $(1+b_{\rm
  vel})^{3/2}$ (corresponding  to $\sim 1.85$  in the case  of $b_{\rm
  vel} =0.5$).

The   results   presented  here   also   have  potentially   important
implications for  (strong) gravitational lensing. In both  the NRG and
NRH scenarios one  expects a strong, `external' shear  due to the dark
matter halo, which should leave signatures in the image configurations
and  time delays  of the  lens.  In  fact, this  `external'  shear may
already have  been detected.  As  shown in Keeton, Kochanek  \& Seljak
(1997),   fitting  four-image   lenses  almost   always   requires  an
independent external shear  that is not aligned with  the light of the
lens.  Although this may  reflect a misalignement between the luminous
galaxy and dark  matter halo, in agreement with  the results presented
here,  there  are  alternative   sources  of  external  shear  (nearby
galaxies,  large-scale  structure along  the  line-of-sight) that  may
leave a  similar signal  in the lens  configuration. A  more thorough,
systematic study  of multiply lensed systems is  therefore required to
put constraints  on the  spatial bias of  brightest halo  galaxies. In
fact, strong  gravitational lensing is  probably the only  method that
can  be used  to detect  an  offset between  halo and  galaxy in  {\it
  individual}  systems, and to  discriminate between  the NRH  and NRG
scenarios.

A non-zero $\langle r_{\rm cen}  \rangle$ also impacts on the internal
structure and  dynamics of central galaxies. As  the galaxy oscillates
in the dark matter halo (NRG scenario), or the halo relaxes around the
central  galaxy (NRH scenario),  it is  constantly subjected  to tidal
forces that  may trigger bar-instabilities in  otherwise stable disks,
may cause excessive  heating of the disk, and  may create lopsidedness
(Levine \& Sparke 1998; Noordermeer, Sparke \& Levine 2001).  Detailed
studies have  revealed lopsidedness (either  in the kinematics  or the
photometry) in about half of  all disk galaxies studied (e.g., Richter
\& Sancisi  1994; Zaritsky  \& Rix 1997;  Rudnick \& Rix  1998; Haynes
\etal 1998; Matthews, van Driel  \& Gallagher 1998; Swaters 1999).  In
fact,  as  shown  by   Bissantz,  Englmaier  \&  Gerhard  (2003),  the
morphology and kinematics of gas in the inner few kpc of the Milky Way
(in  particular the  $3  \kpc$-arm)  may indicate  the  presence of  a
similar $m=1$ asymmetry  in our own galaxy (cf.   Fux 1999).  The high
frequency of lopsided  and barred disk galaxies therefore  seems to be
in support  of a  non-zero $\langle r_{\rm  cen} \rangle$,  whether it
reflects  a  non-relaxed halo  or  a  non-relaxed  galaxy. Taking  our
results at  face value, it is  clear that any study  of disk stability
that  ignores these  strong  distortions and  time-variability of  the
potential may be missing an essential ingredient.

Finally, as  mentioned in the  introduction, a non-zero  $b_{\rm vel}$
also  plays  a  role  in  halo occupation  models.  Using  the  method
described in  detail in Yang  \etal (2004), we computed  the projected
two-point   correlation  function   and  pairwise   peculiar  velocity
dispersions  of   MGRSs  $M_{0.0}$,  $M_{0.5}$,   and  $M_{1.0}$.  The
differences are found to be extremely small, well below the errors due
to  cosmic  variance.   Therefore,  for  all  practical  purposes,  it
suffices  to model  the  phase-space parameters  of  galaxies in  dark
matter  haloes  with $b_{\rm  vel}=0$  (as  is  generally done),  when
computing galaxy-galaxy correlation functions based on halo occupation
distributions.

\section{Conclusions}
\label{sec:concl}

According  to  the  standard  paradigm  of  structure  formation,  the
brightest galaxy  in a dark matter  halo should reside at  rest at the
center of  the potential well. In  order to test  this `central galaxy
paradigm' (CGP),  we used the  halo-based galaxy group finder  of Yang
\etal (2005a) to construct group  catalogues from the 2dFGRS and SDSS. 
For each  group we  compute the statistic  ${\cal R}$, defined  as the
difference between the velocity of  the brightest group galaxy and the
average velocity  of the other group  members (satellites), normalized
by the unbiased estimator of  the velocity dispersion of the satellite
galaxies.  If  the null-hypothesis of  the CGP is correct,  ${\cal R}$
should  follow a  Student $t$-distribution.   If, on  the  other hand,
brightest halo galaxies have a  non-zero velocity bias with respect to
the  satellite   galaxies,  the  ${\cal   R}$-distribution  should  be
significantly  broader.  The applicability  of this  `${\cal R}$-test'
depends  critically on  how well  one  can group  those galaxies  that
belong to the same dark matter halo. Although our group finder is well
tested  and calibrated,  it is  not perfect,  and  unavoidably selects
interloper galaxies  as group  members. In addition,  redshift surveys
suffer from various incompleteness  effects.  We have shown that these
effects result in a  broadening of the ${\cal R}$-distribution, which,
when not accounted for, may give  the false impression that the CGP is
ruled out.

In order to take  interlopers and incompleteness effects into account,
and  thus allow  for a  fair comparison  with the  data,  we construct
detailed mock  galaxy redshift surveys  that can be compared  with the
2dFGRS on a one-to-one basis. We apply our ${\cal R}$-statistic to the
galaxy  groups selected  from these  MGRS, which  we compare  to those
obtained  from the  2dFGRS  using the  Kolmogorov-Smirnov test.   This
shows that the  CGP is inconsistent with the  data at high confidence,
and that instead  the brightest halo galaxies have  a specific kinetic
energy that is about 25 percent  of that of satellite galaxies.  For a
typical,  relaxed,  cold  dark  matter  halo this  corresponds  to  an
expectation value for  the offset between galaxy and  halo of $\sim 3$
percent of the virial  radius, comparable to the characteristic radius
of the  galaxy.  In addition,  we find a  weak hint that  the velocity
bias of brightest halo galaxies is larger in more massive haloes.

We have focussed mainly  on the ${\cal R}$-distributions obtained from
the 2dFGRS, simply  because we have accurate MGRSs  available for this
data  set. However, we  have shown  that the  ${\cal R}$-distributions
obtained from groups in the SDSS are in excellent agreement with those
obtained  from   the  2dFGRS,  suggesting   that  the  SDSS   is  also
inconsistent with the CGP.
 
Undoubtedly,  the most  important implication  of our  results  is the
puzzling question as  to the origin of the  offset between the central
galaxy and  its satellites.  Probably  the most likely  explanation is
that the majority of dark matter  haloes are not yet fully relaxed. In
this case, the brightest halo  galaxy may still coincide with the {\it
  minimum} of the  potential well, but that minimum  does not coincide
with the center  of mass measured over the  entire halo. Although this
picture seems  consistent with our  finding that the  specific kinetic
energy  of the  brightest  halo  galaxies is  larger  in more  massive
haloes, which form later,  detailed numerical simulations are required
to investigate whether the typical  growth rates of dark matter haloes
are  sufficiently large  and  violent  to explain  our  findings in  a
$\Lambda$CDM concordance  cosmology. A first hint that  this is indeed
the  case comes  from the  simulations  of Diaferio  \etal (1999)  and
Yoshikawa \etal (2003), which reveal a velocity bias of brightest halo
galaxies  that is  very similar  to  that found  here.  Although  this
suggests  that  a non-zero  velocity  bias  is  a natural  outcome  of
structure formation in a $\Lambda$CDM  cosmology, it still needs to be
verified whether, in these simulations,  the central galaxy is at rest
with respect to the {\it minimum} of the potential well.

An alternative explanation for  the non-zero velocity of the brightest
halo galaxies with  respect to the satellite galaxies  may be that the
halo is relaxed, but that  the brightest halo galaxy oscillates in the
central  potential  well.   If   the  dark  matter  halo  is  strongly
concentrated,  as expected for  typical cold  dark matter  haloes, one
would naively  expect that any  such oscillation is quickly  damped by
dynamical   friction.   However,  this   damping   timescale  may   be
significantly longer if there is less dark matter in the center of the
halo than anticipated; i.e.,  the density distribution is cored rather
than cusped.  This possibility is  interesting in light of  the recent
claims for cored haloes based  on the observed rotation curves and bar
pattern speeds of disk galaxies.

In either case, the brightest  halo galaxy is expected to experience a
time-varying tidal field. This strongly questions the applicability of
(numerical) studies of galaxy dynamics, and in particular of stability
analyzes that  make the assumption that  the galaxy is at  rest at the
center of  a relaxed dark matter  halo. In particular,  it may explain
the  high frequency  and longevity  of bars  and lopsidedness  in disk
galaxies. The  fact that we find  evidence for a  non-zero velocity of
the brightest halo galaxy with  respect to the satellite galaxies also
has important implications for  the determination of halo masses based
on the kinematics  of host-satellite systems, and for  the modeling of
strong   gravitational   lenses.   For   the   purpose  of   computing
galaxy-galaxy correlation  functions based on  halo occupation models,
however, one  can safely ignore the  fact that the CGP  does not hold,
and make the  simple ansatz that the brightest  halo galaxy resides at
rest at the halo center.


\section*{Acknowledgements}

We are grateful to Michael Blanton for his help with the NYU-VAGC, and
to Victor Debattista, Savvas  Koushiappas, George Lake, Shude Mao, Ben
Moore,  Peter Schneider, Joachim  Stadel, and  Simon White  for useful
discussions.



\label{lastpage}

\end{document}